\documentclass[sigconf]{acmart}
\usepackage{graphicx}
\usepackage{comment}

\usepackage{multirow}
\usepackage{textcomp}
\usepackage{subcaption}
\AtBeginDocument{%
  }

\setcopyright{none}

\acmConference[ASP-DAC '27]{Tokyo}{Japan}

\begin{document}

\title{A Time-Encoded Analog Photonic Interposer for Energy-Efficient Integration of Analog Vision Sensors and Analog Accelerators}

\author{%
  Subhradip Chakraborty$^1$,
  Zihan Yin$^1$,
  Xuming Chen$^2$,
  Chengwei Zhou$^2$,
  Gourav Datta$^2$,
  and Akhilesh Jaiswal$^1$
}

\affiliation{%
  \institution{$^1$University of Wisconsin--Madison, $^2$Case Western Reserve University}
  \country{Madison, WI, USA}
}

\email{chakrabort42@wisc.edu,  akhilesh.jaiswal@wisc.edu}

\begin{abstract}

This work introduces a time-encoded analog photonic interposer that enables long-distance, high-fidelity transport of analog signals between spatially separated chiplets. Unlike prior silicon-photonic links limited to digital data, the interposer preserves analog information by converting amplitudes into timing intervals using an analog-to-time converter (ATC), transmitting them over a wavelength-division-multiplexed photonic link, and reconstructing values at the receiver without an explicit high-precision ADC/DAC data-converter pipeline. The link instead embeds an \emph{implicit} 6-bit time-domain quantization and uses a single wavelength per processing element independent of bit precision. Evaluated in a fully analog vision pipeline with an in-pixel computing (IPC) sensor, it achieves a 2.04$\times$ energy--delay product (EDP) improvement over an 8-bit digital electrical baseline on the 560$\times$560 Visual Wake Words dataset, with the advantage widening with link length even against a precision-matched 6-bit baseline. The pipeline holds 89.87\% and 86.15\% accuracy on ResNet18 and MobileNetV2 for VWW and generalizes across CIFAR-10 and ModelNet40 within 2\% of the digital baseline.



\end{abstract}


\maketitle

\section{Introduction}

The growing demand for distributed, multi-chip edge AI systems has placed renewed emphasis on the communication bottleneck between physically separated sensor front-ends and downstream neural network accelerators~\cite{ia,ib,ic,id}. While in-sensor analog compute techniques such as Processing-in-Pixel Computing (IPC)~\cite{datta2022processing, ipc1} significantly reduce data volume at the front-end, their benefits are ultimately limited by the need to export analog activations across chiplets~\cite{multchip1, mulchip2}. Existing systems overwhelmingly rely on electrical interposers for this task, often necessitating ADC/DAC conversion~\cite{cha1, cha2, cha3}. As the communication distance grows, electrical signaling suffers from resistive loss, capacitive loading, and bandwidth limitations~\cite{elec1, elec2, electricalchallenge}, requiring repeated digitization or buffering stages that erode the advantages of analog-domain computation. The challenge is therefore not only to compute efficiently within the sensor plane but also to preserve analog information across physically distributed units. To this end, photonic interposers offer high-speed inter-chip transport~\cite{opticalinterconnect1, opticalinterconnect2}: silicon-photonic waveguides and microring resonators (MRRs) provide extremely high bandwidth and minimal parasitics, scaling far better with distance than electrical links~\cite{opticalinterconnect3, opticalinterconnect4}. However, prior optical interposers support only digital data transfer and are unsuitable for analog signals.

\begin{figure}[!t]
\centering
\includegraphics[width=0.9\linewidth]{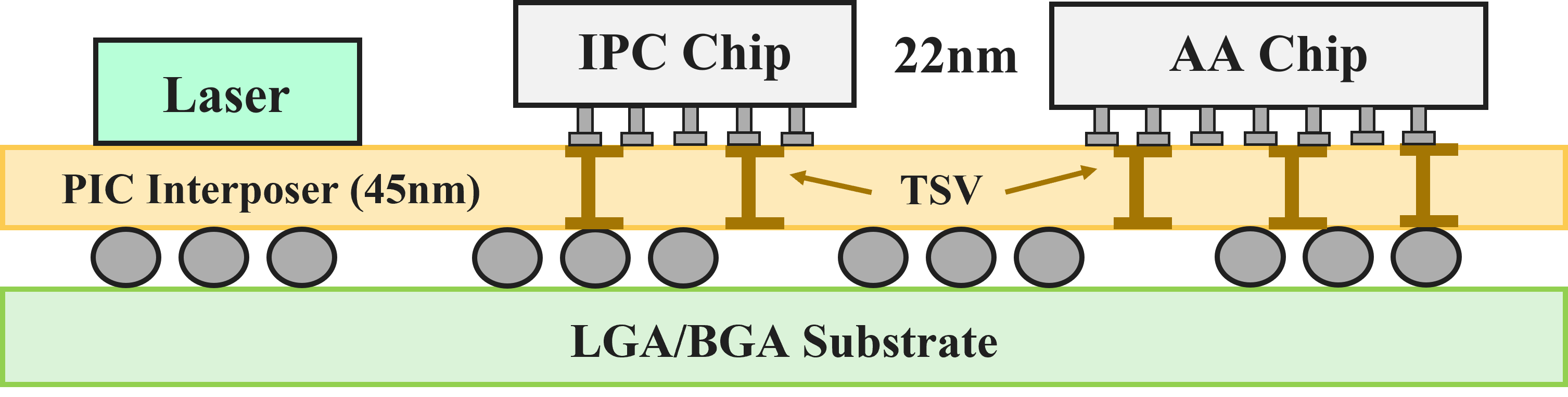}
\vspace{-0.1 in}
\caption{Overview of the photonic interposer connecting the IPC sensor with analog accelerator (AA).}

\vspace{-0.1 in}
\label{PIC}
\Description{IPC arch}
\end{figure}

\begin{figure*}[!ht]
\centering
\centerline{\includegraphics[width=0.82\textwidth]{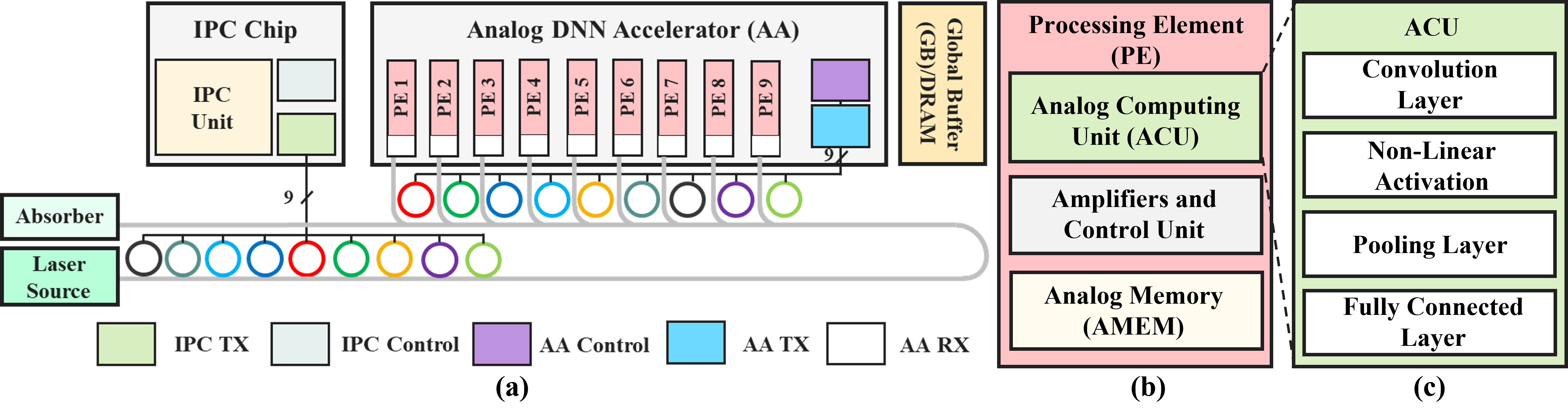}}
\vspace{-0.2 in}
\caption{(a) Proposed architecture diagram; (b) Processing element (PE) within the analog accelerator, consisting of the computing unit, analog memory, and control unit; (c) Analog computing unit responsible for convolution, non-linear activation, pooling, and fully connected operations.}
\Description{Architecture diagram of the system.}
\vspace{-0.1 in}
\label{fig1_Analog_Interconnect_Overview}
\end{figure*}

This limitations motivates the need for an interposer capable of transporting analog signals seamlessly, preserving analog information without requiring an explicit high-precision ADC/DAC data-converter pipeline as shown in Fig.~\ref{PIC} and~\ref{fig1_Analog_Interconnect_Overview}. In this work, we introduce a time-encoded analog photonic interposer, which performs analog-to-time conversion (ATC) to map analog voltages into timing intervals that can be robustly transmitted over wavelength-division-multiplexed (WDM) optical channels. At the receiver end, the timing information is reconstructed back into analog voltages using a complementary time-to-analog mechanism. Time-domain encoding circumvents the limited MRR extinction ratio (15–25 dB~\cite{mrrext1, mrrext2}) by mapping information onto timing variations rather than amplitude, enabling robust transfer of direct analog signals. While this interposer can interface with any high throughput analog compute front-end, we demonstrate its utility using a fully analog image-processing pipeline: an IPC sensor that produces first-layer analog activations and an analog accelerator that executes deeper network layers. This 2.5D heterogeneous system highlights how analog information can be generated, stored, and transmitted. The main contributions of this work are:

\begin{itemize}
    \item We propose a novel time-encoded analog photonic interposer that enables long-distance, energy-efficient transport of analog signals across a silicon-photonic interposer. Time-encoding maps information to timing instead of amplitude, overcoming MRR limitations and enabling robust analog transfer.
    \item We demonstrate the use of proposed interposer to build a fully analog sensor-to-processor path that avoids an explicit high-precision ADC/DAC data-converter pipeline between chiplets, instead embedding an implicit 6-bit time-domain quantization within the link itself.
    \item We integrate the interposer within a realistic vision pipeline
    using an IPC front-end and an analog accelerator, enabling
    end-to-end evaluation on a 560$\times$560 VWW-dataset~\cite{vww}.
    \item Hardware simulations on commercial GF~22nm for electronic chiplets and GF~45SPCLO technology for photonic interposer  show a 2.04$\times$ EDP improvement over electrical interconnects for long-distance communication.
\end{itemize}


\section{Computing Preliminaries}
\subsection{IGZO Based Gain Cell}
 Indium gallium zinc oxide (IGZO) devices~\cite{9401719,9212585,9815041} offer very high resistance and low leakage, enabling long retention without refresh and without large metal oxide--metal (MOM) capacitors~\cite{shi2018evolution}. Their compact footprint allows dense analog memory across the pixel plane for large arrays such as our $560\times560$ VWW configuration. We use their near-nonvolatile behavior as an analog buffer that holds stored activations while they are transmitted over the photonic link.

\subsection{Micro Ring Resonator (MRR) \& Thermal Calibration}

\begin{figure}[!t]
\centering
\includegraphics[width=1\linewidth]{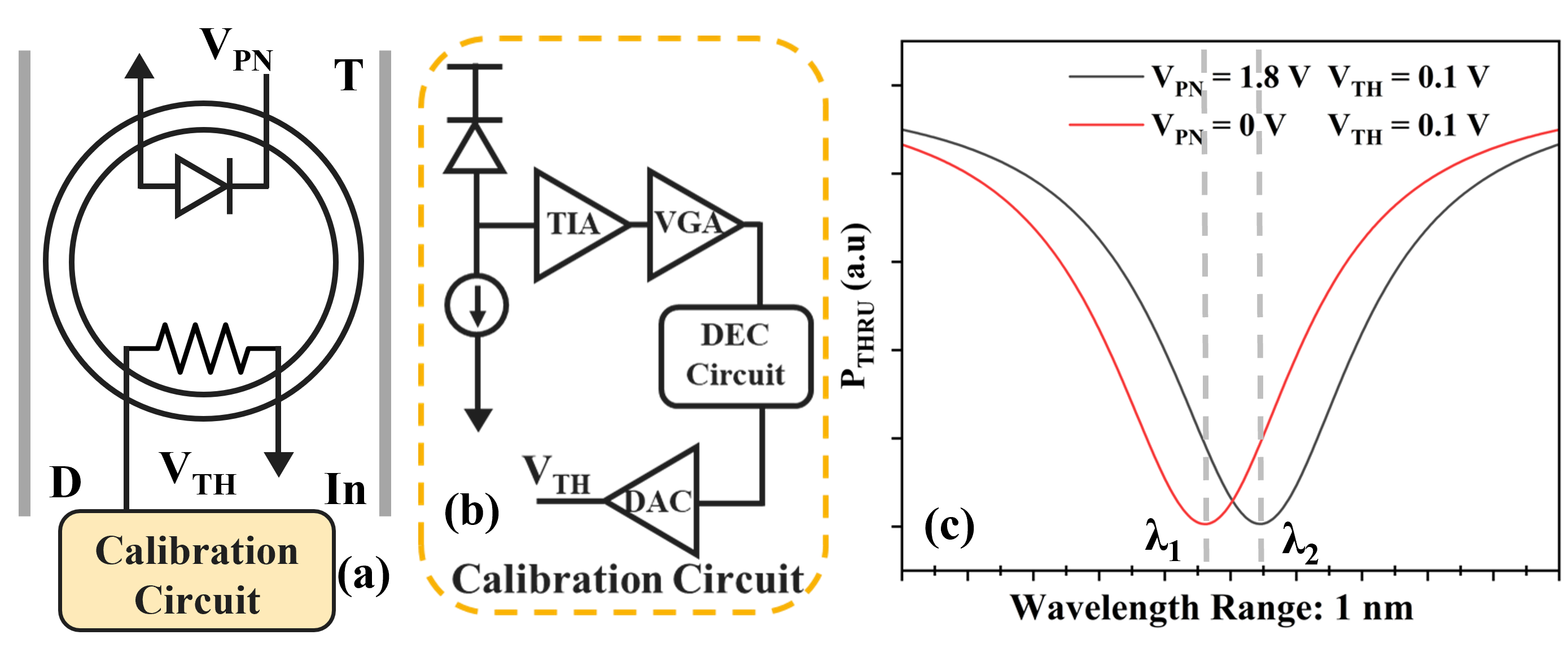}
\vspace{-0.2 in}
\caption{(a) 4-port MRR; (b) thermal tuning calibration circuit; (c) through port transmission spectrum vs. wavelength.}
\Description{aa}
\vspace{-0.1 in}
\label{MRR_Intro}
\end{figure}

Fig.~\ref{MRR_Intro}(c) shows the transmission of a four-port MRR~\cite{mrrintro} on the
\\GF45SPCLO, where the resonance condition sets the through (T) port power. Resonance is tuned by biasing the integrated pn-junction (n-terminal swept, p grounded, thermal-tuning voltage fixed): for a fixed wavelength, $V_{PN}=1.8$ shifts resonance so less power reaches the through port, while $V_{PN}=0$ transmits more. Because mismatch, process variation, and thermal drift shift the resonance~\cite{mrrthermal}, calibration is required; prior analog/mixed-signal schemes~\cite{thermalcal1, thermalcal2} (Fig.~\ref{MRR_Intro}(b)) sense the drop (D) port with a photodiode and drive the thermal port ($V_{TH}$) through a DAC in closed loop.

 \begin{figure}[!t]
\centering
\includegraphics[width=0.9\linewidth]{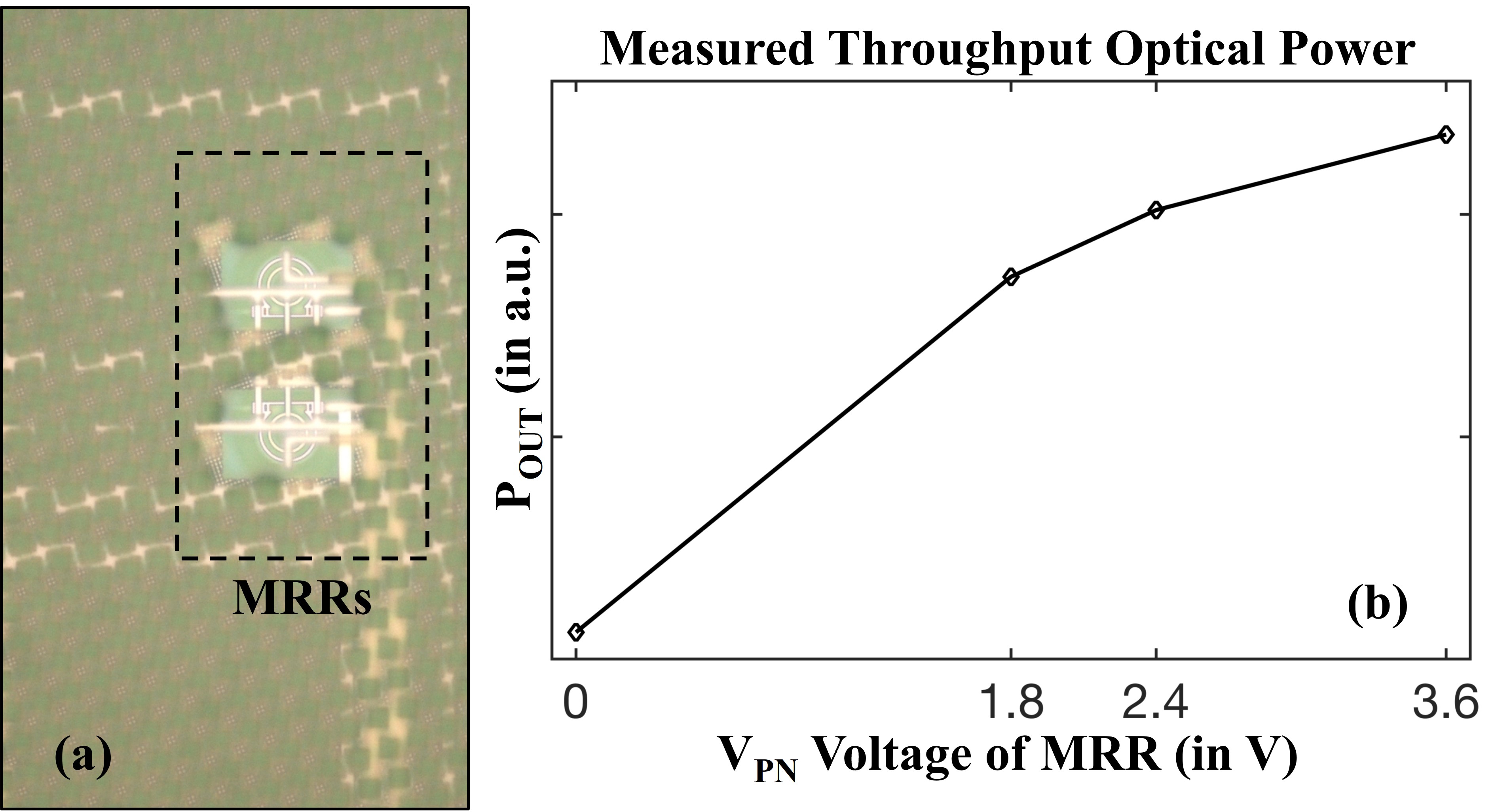}
\vspace{-0.1 in}
\caption{(a) Chip micrograph with a zoomed-in view of two MRR configuration fabricated using GF45SPCLO technology; (b) Measured through port output response of a 4-port MRR against different PN voltages.}
\vspace{-0.1 in}
\label{MRR_Measurement}
\end{figure}

Fig.~\ref{MRR_Measurement} shows the silicon measurements results of a 4-port MRR in GF45SPCLO. Varying $V_{PN}$ modulates the light coupled into the ring and thus the through-port power; while higher $V_{PN}$ raises the power, the margin between levels shrinks at high bias and large swings are needed, making direct analog transmission difficult. We therefore encode analog information in the time domain using only two voltage levels, preserving margin and avoiding high-voltage interfacing.

\section{Proposed Architecture}
\subsection{Overview}

Fig.~\ref{fig1_Analog_Interconnect_Overview} shows the proposed system, centered on a time-encoded analog photonic interposer for energy-efficient, long-distance analog transport between chiplets. We demonstrate it with a fully analog vision pipeline: (1) an IPC image sensor performing the first convolution and outputting analog activations, (2) the photonic interconnect on a silicon interposer, and (3) an analog accelerator (AA) executing the remaining layers following prior analog DNN designs~\cite{10299653,9731773,10689660}. The interposer is our focus; we adopt the accelerator of~\cite{10299653} and concentrate on the IPC front-end and the time-encoded interposer that enables inter-chip communication without a conventional ADC/DAC pipeline.

\begin{figure}[!t]
\centering
\includegraphics[width=1\linewidth]{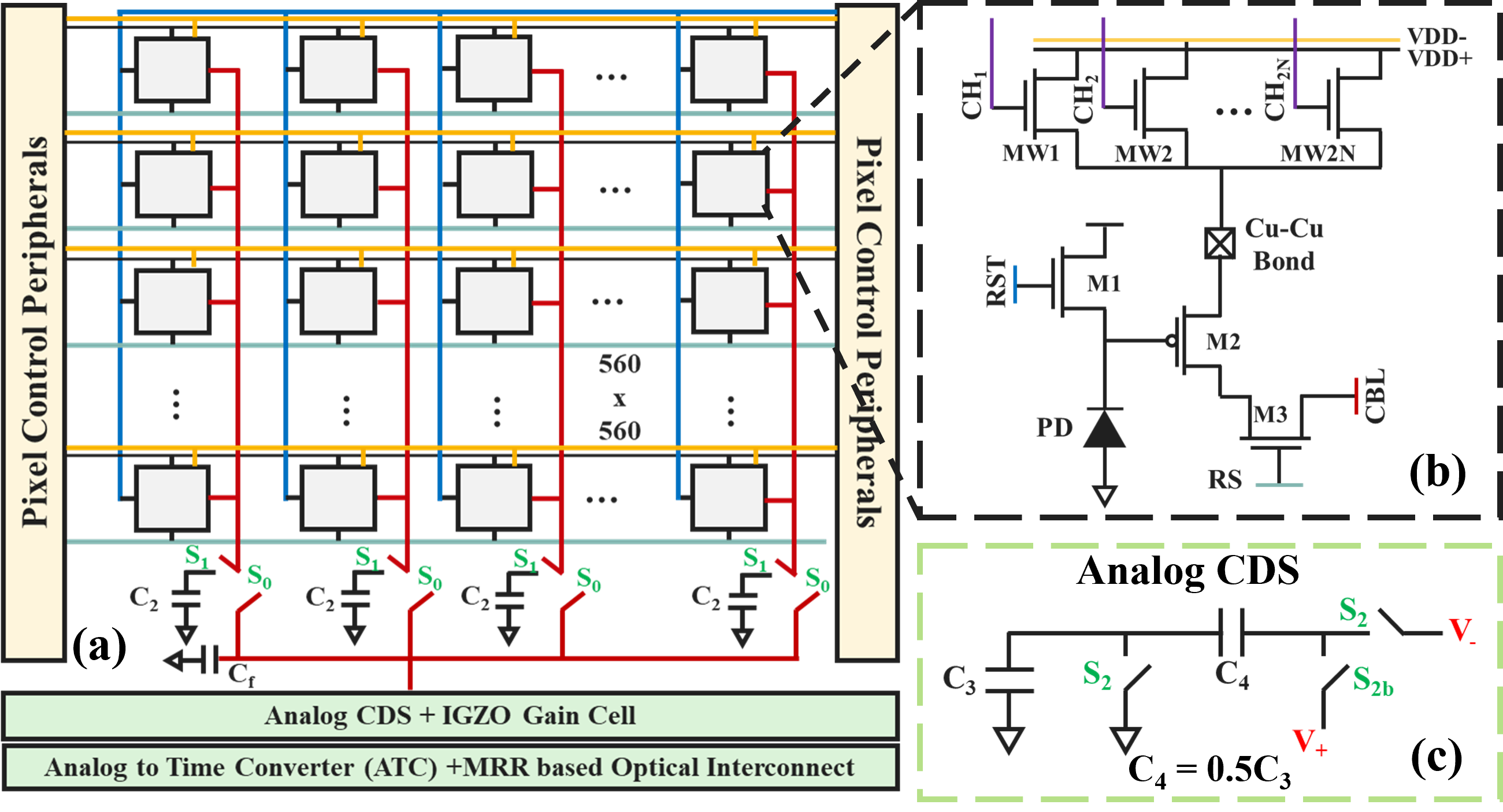}
\vspace{-0.2 in}
\caption{Architecture of the IPC chip (a) P2M pixel array with peripheral CDS and ATC units, (b) unit pixel circuit, and (c) analog CDS design.}
\vspace{-0.1 in}
\label{fig2_IPC_Array}
\Description{IPC arch}
\end{figure}

\subsection{In-Pixel Computing Sensor}

The IPC chip serves as the analog front-end, providing the first-layer activations transmitted through the interposer. We modify the Processing-in-Pixel-in-Memory (P2M) unit of~\cite{datta2022processing} into a fully analog architecture as shown in Fig.~\ref{fig2_IPC_Array} that computes the first convolution directly inside the pixel array. The 3T CIS pixels connect to fixed-weight transistors through Cu--Cu hybrid bonds, producing a current proportional to the input--weight product; column outputs are accumulated and averaged on the shared CBL and stored as a voltage on $C_2$, completing the MAC. For $K=3$, the three column voltages are accumulated on $C_f$ and forwarded to the analog correlated double sampling (CDS) unit. To support signed convolution, each weight is split into positive and negative components sampled over two exposure cycles, and the CDS circuit (Fig.~\ref{fig2_IPC_Array}(d)) subtracts them as $(V_{+}-V_{-})/3$, yielding a signed MAC output without digitization. The activations are stored in compact, low-leakage IGZO gain cells whose retention suffices to forward them to the ATC that drives the photonic interposer.

\begin{figure}[!ht]
\centering
\includegraphics[width=1\linewidth]{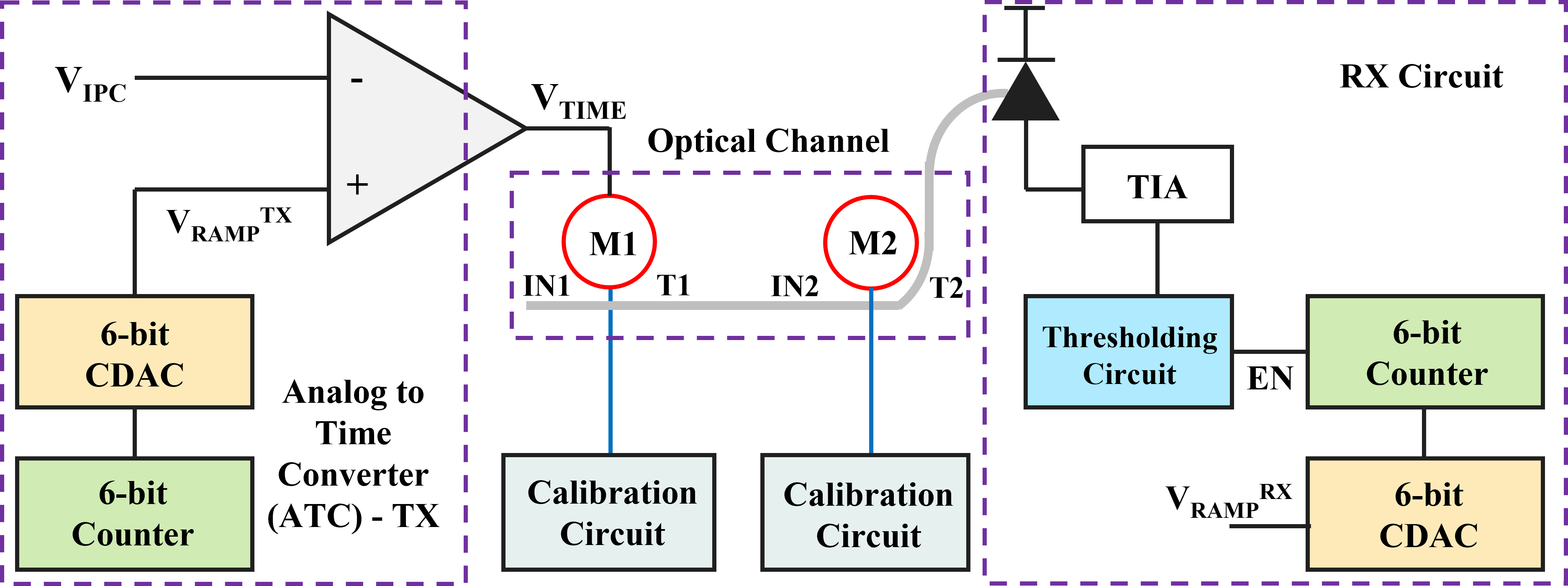}
\vspace{-0.1 in}
\caption{The transmitter (TX) and receiver (RX) circuits along with optical interposer showing only the through port (T) and input port (IN) of the MRRs.}
\vspace{-0.1 in}
\label{TX_RX_Analog}
\end{figure}

\subsection{Analog Photonic Interposer}\label{sec3-photonic}

\subsubsection{Transmitter Design}
The IPC chip's transmitter sends the first-layer analog outputs serially through the optical interposer using an ATC as shown in Fig.~\ref{TX_RX_Analog}. Converting the analog voltage into a time-encoded signal lets MRRs carry the information as a timing edge; an optical link is required since an electrical one would add latency and offset that corrupt the timing. The ATC comprises a 6-bit ramp generator, a 6-bit counter driving a 6-bit CDAC that produces a ramp $V_{RAMP}^{TX}$ each clock, and a strong-arm latched comparator at the same clock. When the ramp exceeds $V_{IPC}$, the comparator output ($V_{TIME}$) switches $1\!\to\!0$ and, after an inverter, drives the $V_{PN}$ of MRR M1: $V_{TIME}=0$ gives high through-port power while $V_{TIME}=1$ lowers it. This maps the analog voltage to a time-encoded optical signal for direct transmission.

\subsubsection{Receiver Design}

The receiver sits within each PE of the AA chip and receives the optical input directly, without global buffering. The incoming light couples into MRR M2 (held at $V_{PN}=0$, off-resonance), whose through port feeds a photodiode; a transimpedance amplifier (TIA) converts the photocurrent to voltage, and a thresholding stage outputs logic 1 when the TX sends high optical power ($V_{PN}=0$ at TX) and 0 otherwise. This output as an enable signal for the 6-bit RX ramp generator ($V_{RAMP}^{RX}$). The ramp runs while EN is high and, when EN falls, halts and samples its analog output onto a capacitor in the PE, reconstructing the transmitted value and completing analog recovery across the interposer.

\subsubsection{Relationship to Conventional Data Converters}
Although the transmitter shares building blocks with a single-slope ADC front-end (counter, CDAC, comparator), the link is better described as a time-domain analog interconnect with \emph{implicit} 6-bit quantization rather than an explicit ADC/DAC pipeline. No digital codeword is generated or transmitted: each value is a single optical timing edge, eliminating the capture registers, encoders, and serializer/deserializer pair a digital readout requires. This also minimizes the wavelength budget, a 6-bit digital photonic link would need six wavelengths per value, scaling with precision and kernel size until it exceeds the FSR of the GF45SPCLO platform, whereas our scheme uses one wavelength per PE independent of precision, at the cost of a longer, data-dependent pulse hidden by wavelength-parallel transmission. Finally, since the receiver samples an analog quantity derived from the edge, a timing error yields only a small voltage deviation rather than the catastrophic bit flip of an NRZ link, and the RX thresholding stage can fold application-specific nonlinearities (e.g., activations) into recovery, providing in-link computation unavailable to a conventional DAC receiver.

\begin{figure}[!ht]
\centering
\includegraphics[width=1\linewidth]{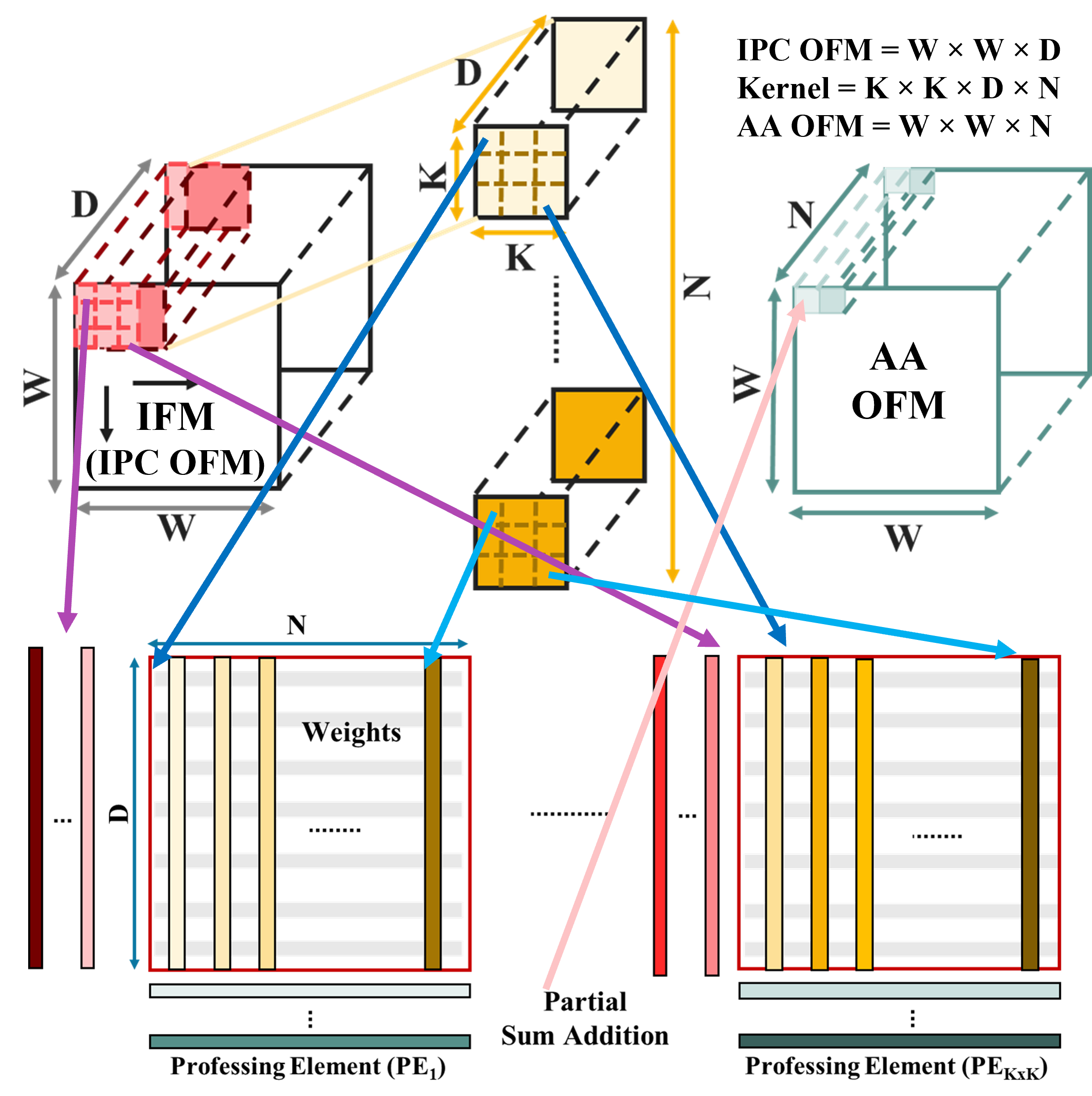}
\vspace{-0.2 in}
\caption{The IPC-generated OFM is forwarded to the AA, where it is mapped and processed for the next layer.}
\vspace{-0.2 in}
\label{fig-Mapping}
\end{figure}

\subsection{Mapping of a neural network}

The interposer uses a weight-stationary mapping that reduces data movement~\cite{mapping} and exploits MRR wavelength selectivity. Since the first layer runs on the IPC chip, we describe how its outputs map onto the AA for subsequent layers. The AA activates $K\times K$ PEs for the second-layer convolution, each with an MRR tuned to a unique wavelength $\lambda_1\cdots\lambda_{K\times K}$ (e.g., 9 PEs/wavelengths for $K=3$). Since each MRR couples power only at its wavelength, the interposer performs spatial demultiplexing without electronic routing.

As shown in Fig.~\ref{fig-Mapping}, PE$_1$ stores the first coefficient of all $N$ output channels, PE$_2$ the second, and so on (stride 1), turning the computation into a streamed activation flow in which the second-layer IFM values (the IPC's OFM activations) are transferred from the IPC to the AA. These values are sent across the interposer on multiple wavelengths simultaneously: sample $i$ is encoded on $\lambda_i$ and received only by PE$_i$, repeating cyclically until every PE holds the activation to multiply with its stationary weight; the partial products are then accumulated locally in analog form.

For deeper layers, IFMs are generated inside the AA and processed locally, with weights loaded from a global memory (e.g., DRAM) via optical or electrical interconnect~\cite{opticalinterconnect3}, following the conventional DNN mapping with more PEs activated for larger kernels. Because the time-encoded link commits a single wavelength per PE independent of bit precision, the WDM channel count scales with kernel size ($K\times K$) rather than precision; the $K=3$ second-layer transfer uses nine wavelengths within the GF45SPCLO FSR, while larger kernels activate proportionally more wavelengths up to the FSR bound. The time-encoded WDM link is thus dedicated to the long-distance sensor-to-accelerator activation transfer where its efficiency advantage is greatest, streaming only activations while weights remain stationary.

\section{Results and Discussion}

\begin{figure}[!t]
\centering
\includegraphics[width=1\linewidth]{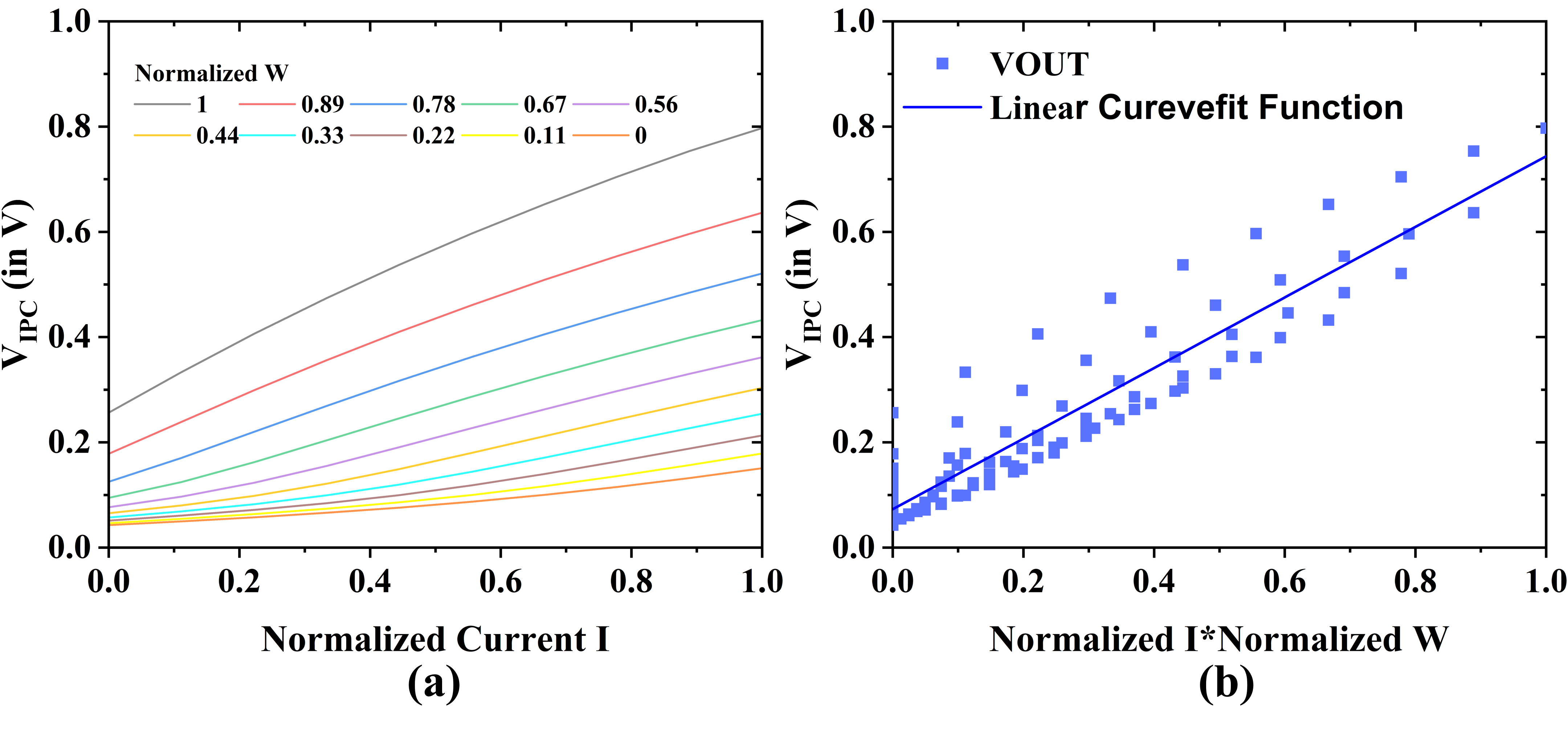}
\vspace{-0.3 in}
\caption{Linearity analysis of the IPC output voltage with respect to (a) normalized input current \(I\), and (b) normalized MAC operation \((I \times W)\).}
\Description{aa}
\vspace{-0.1 in}
\label{IPC_Linearity}
\end{figure}

\begin{figure}[!t]
\centering
\includegraphics[width=1\linewidth]{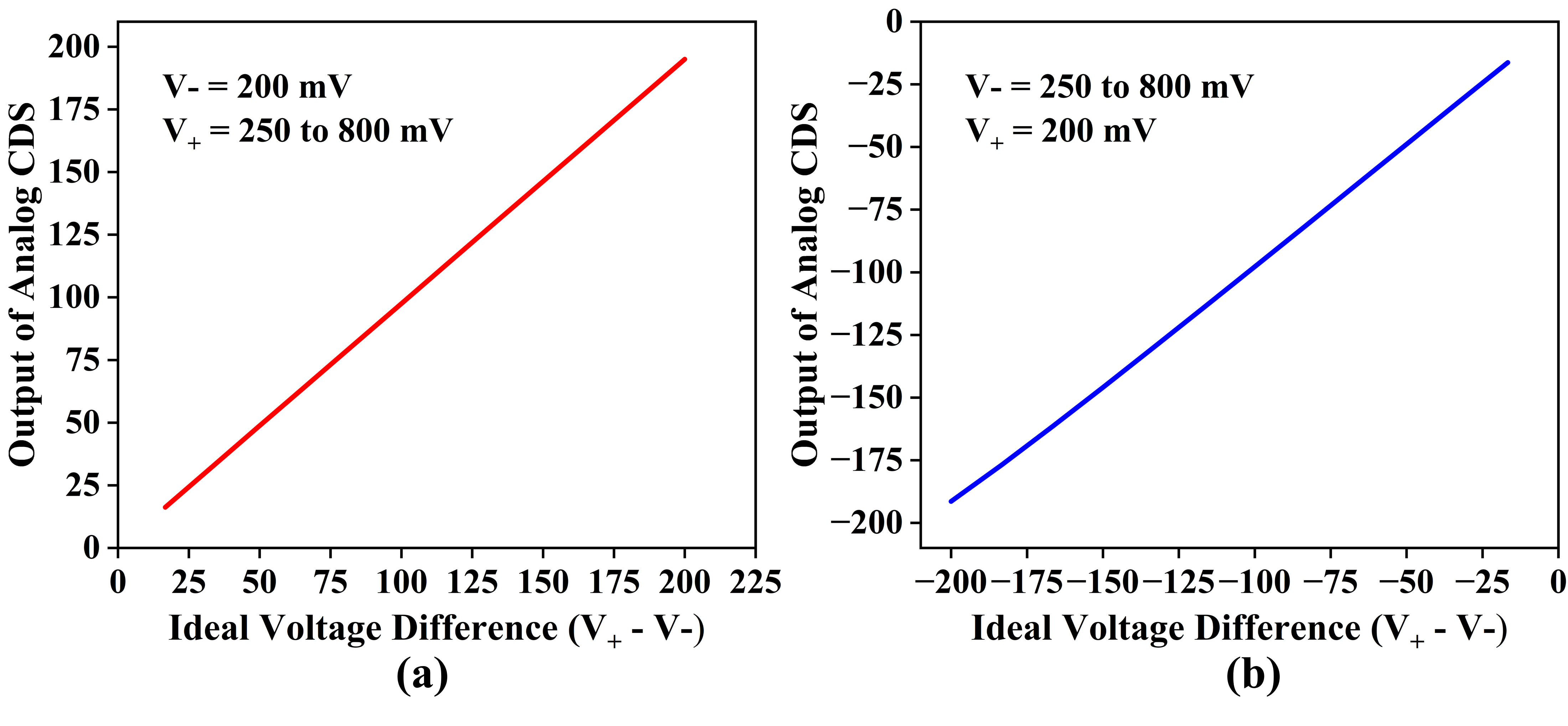}
\vspace{-0.3 in}
\caption{The signed MAC computation for (a) positive and (b) negative cases computed by analog CDS.}
\Description{aa}
\vspace{-0.1 in}
\label{CDS}
\end{figure}

\subsection{Hardware Performance}


\begin{figure}[!t]
\centering
\includegraphics[width=1\linewidth]{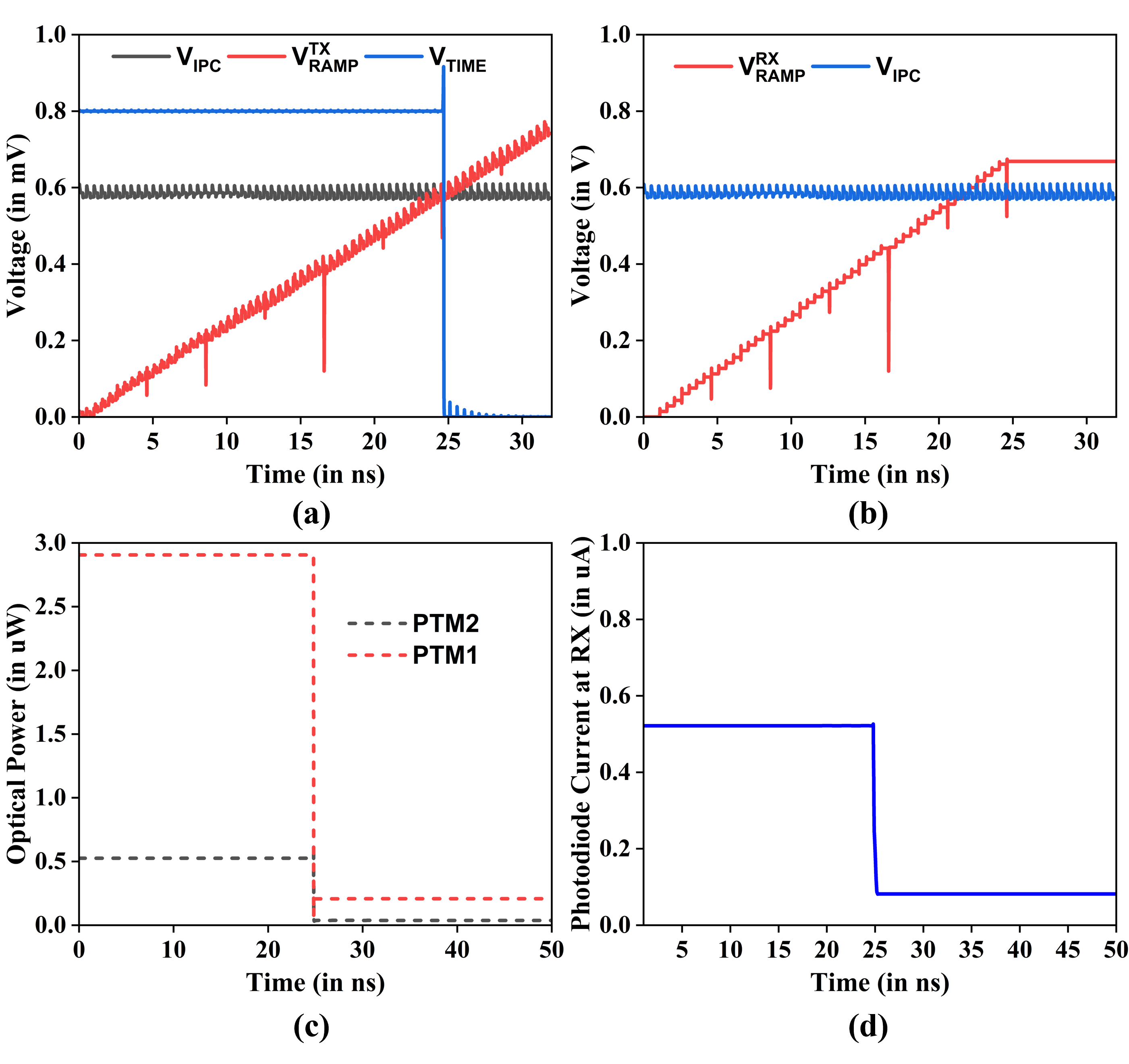}
\vspace{-0.3 in}
\caption{Transient analysis: (a) voltage at the TX end; (b) voltage at the RX end; 
(c) through-port optical power; and (d) photodiode current at the RX.
}
\Description{aa}
\vspace{-0.1 in}
\label{ATC_Transient}
\end{figure}

\begin{figure}[!t]
\centering
\includegraphics[width=1\linewidth]{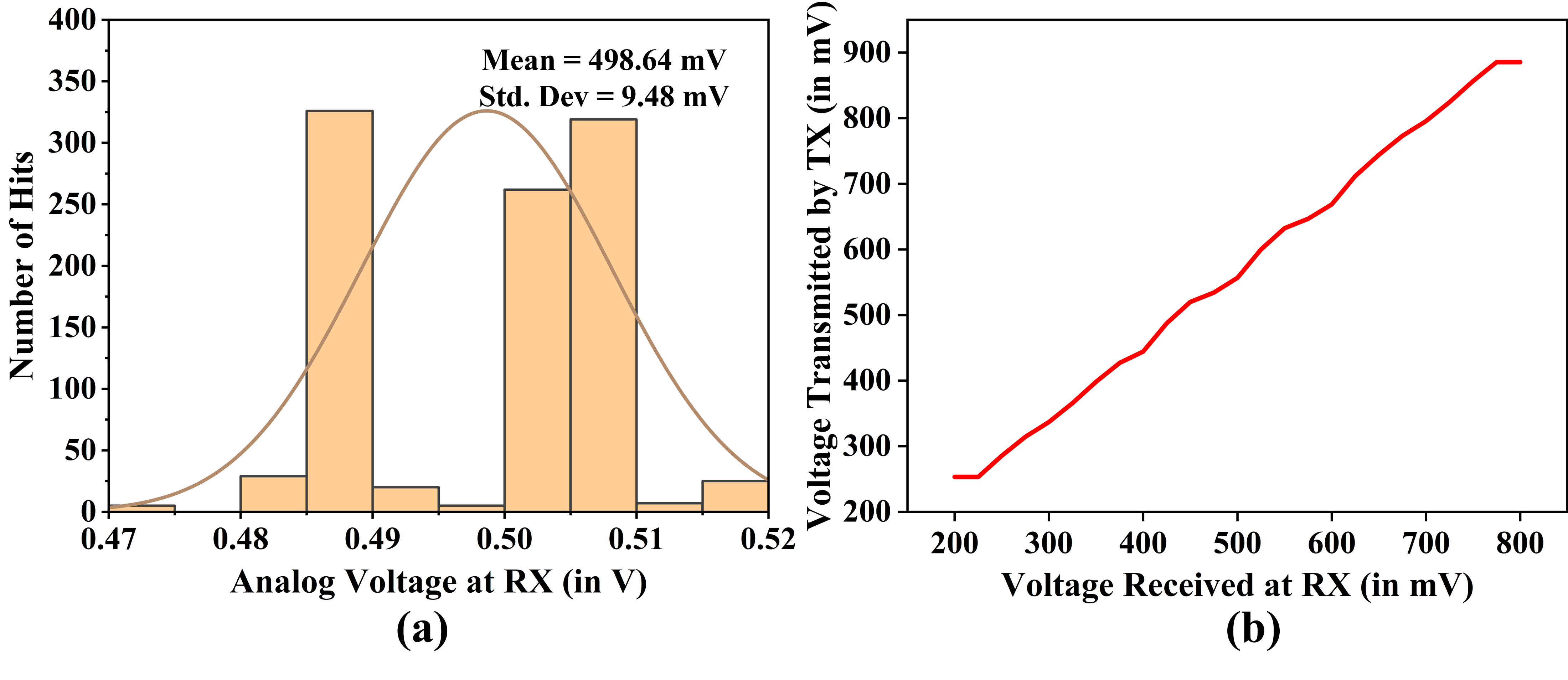}
\vspace{-0.3 in}
\caption{(a) Variation of the RX output voltage when a 450\,mV signal is transmitted from the TX; 
(b) linearity analysis of analog voltage transmission using the proposed architecture.
}
\Description{aa}
\vspace{-0.1 in}
\label{Linearity_Interconnect}
\end{figure}

\subsubsection{Design Analysis}

An end-to-end simulation is performed using the GF PDKs for both optical and electrical components. The linearity analysis of the IPC unit, which computes the first layer of the neural network, is shown in Fig.~\ref{IPC_Linearity}. The analog CDS which performed the signed computation, the deviation from the ideal subtraction value is shown in Fig.~\ref{CDS}. The analog convolution output is stored in an IGZO-based gain cell and subsequently forwarded to the ATC module. Fig.~\ref{ATC_Transient} presents the transient analysis for transmitting an analog value equivalent to 600 mV from the TX on the IPC chip to the RX on the AA chip. The results indicate a bias in the received output voltage at the RX, primarily attributed to kickback noise and comparator offset. Fig.~\ref{ATC_Transient}(c) illustrates the optical time-encoded signal by showing the output power at the through port of the TX-side MRR (PTM1) and the RX-side MRR (PTM2). The corresponding photocurrent at the input of the TIA is shown in Fig.~\ref{ATC_Transient}(d), where currents above 0.2 µA are interpreted as logic ‘1’, and lower values as logic ‘0’. A Monte Carlo analysis with 1000 samples is performed to evaluate the robustness of TX–RX analog data transfer. As shown in Fig.~\ref{Linearity_Interconnect}(a), the deviation in the received voltage for a sample input of 450 mV is minimal, demonstrating strong linearity and stability. Fig.~\ref{Linearity_Interconnect}(b) further illustrates the TX–RX transfer characteristics across a range of input analog voltages. Prior analog-over-photonics works~\cite{opticalinterconnect3, prioropt} rely on a large number of wavelengths, which is impractical for commercial MRRs whose optical response repeats beyond the full spectral range (FSR)~\cite{mrr2}; our design instead uses up to nine wavelengths well within the FSR to transmit data from the IPC chip to the AA chip for the second layer. Moreover, the digital photonic interconnects of~\cite{opticalinterconnect3, opticalinterconnect4} target PE-to-PE communication within an accelerator, whereas our link transports analog activations from a sensor to a remote accelerator over long distances, so its objectives, requirements, and metrics differ substantially. MRR calibration is essential for correct operation, since process variation, device mismatch, and thermal drift shift
  the resonance away from its nominal wavelength and, if uncorrected, would degrade the recovered pulse timing. A key
  advantage of the proposed time-encoding is that, although the transported payload is an analog value, it is conveyed
  in an effectively digital form, the information resides in the pulse width using only two optical levels rather than
  in the optical amplitude. This makes the link naturally compatible with the robust closed-loop \emph{digital}
  calibration scheme of~\cite{thermalcal1}, which realigns the resonance without the stringent amplitude precision
  demanded by conventional analog signaling; consequently, the link tolerates larger resonance drift while keeping the
  calibration overhead low. We further emphasize that the proposed architecture is not restricted to digital
  calibration: it supports the analog calibration schemes of~\cite{thermalcal2, mrrthermal}, allowing the
  designer to trade calibration overhead against loop complexity and accuracy depending on the deployment requirements.

\begin{figure}[!t]
\centering
\includegraphics[width=0.9\linewidth]{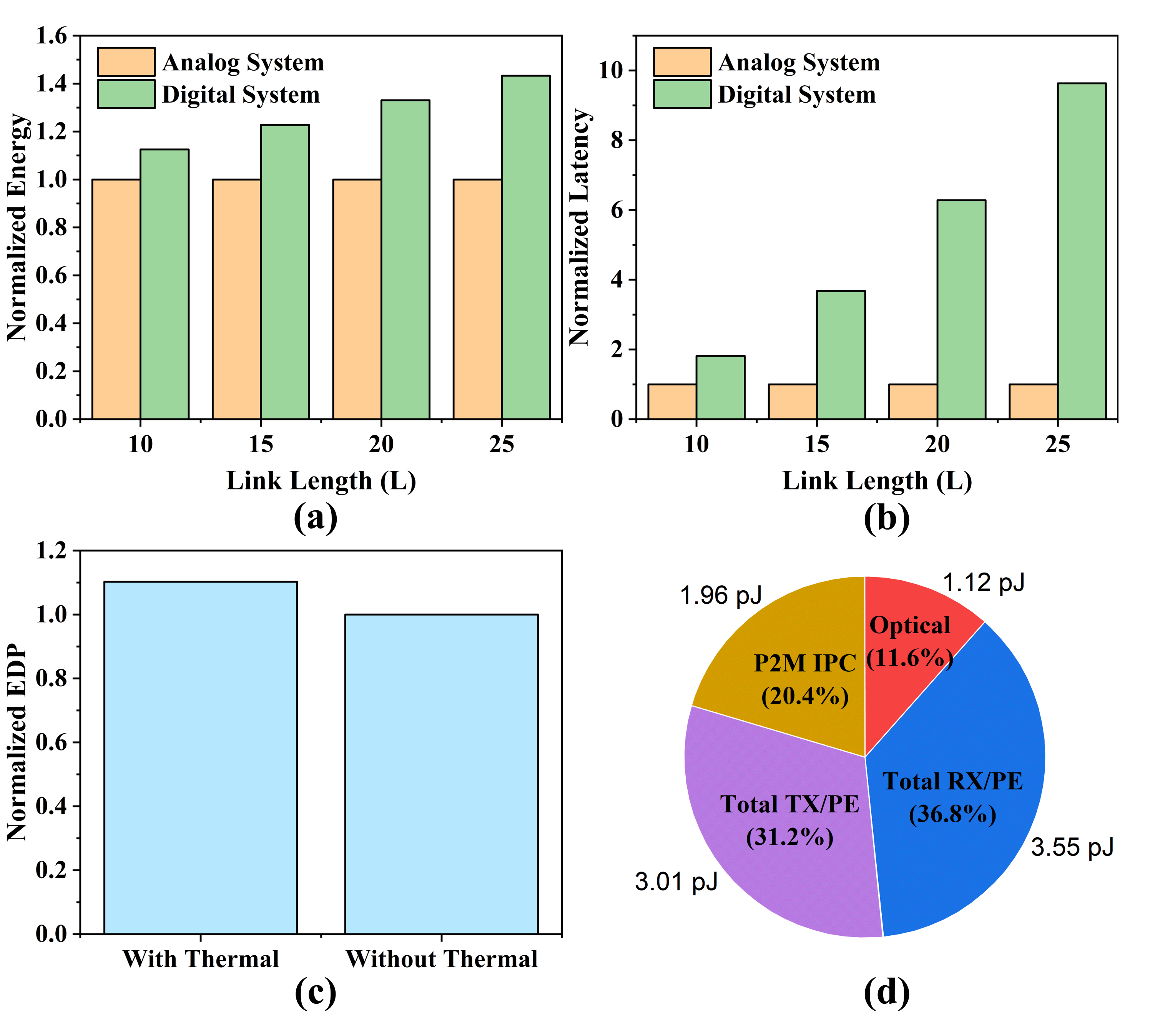}
\vspace{-0.1 in}
\caption{Normalized (a) energy and (b) latency variation with link length for the proposed analog GS IPC system and the digital GS IPC system; (c) Impact of the thermal calibration circuit on the overall EDP; (d) Energy breakdown for individual components of the proposed analog RS IPC design.}
\Description{aa}
\vspace{-0.1 in}
\label{bar_comp}
\end{figure}

Link characterization: The ATC comprises a 6-bit counter, a 6-bit CDAC, and a strong-arm latched comparator clocked at $1\,\mathrm{GHz}$, giving a 64-cycle ($64\,\mathrm{ns}$) conversion and a per-wavelength sampling rate of $15.6\,\mathrm{MS/s}$; the nine parallel wavelengths of the $K=3$ mapping yield an aggregate $\approx 140\,\mathrm{MS/s}$. As the optical channel ($\sim 40\,\mathrm{GHz}$) is far faster than the ATC, the link rate is set by the time-encoding process; the optical edge transitions on a sub-nanosecond scale limited by the comparator and MRR electro-optic bandwidth, and the transmitted pulse width is data-dependent. The closed-loop thermal-calibration path (photodiode, feedback, and DAC on $V_{TH}$) adds $\sim 1\,\mathrm{pJ}$ per value ($\sim 15\%$ link-energy overhead, Fig.~\ref{bar_comp}(c)); between frames the counter and comparator are clock-gated, so idle power reduces to this calibration overhead alone.

\subsubsection{Energy and Latency}

We evaluate an IPC array with $K=5$, non-overlapping stride 5, and 8 output channels, giving a $112\times112\times8$ OFM (one IGZO cell per OFM value). The interconnect consumes $6.6\,\mathrm{pJ}$ ($7.6\,\mathrm{pJ}$ with thermal calibration) per value; each IGZO write costs $4.8\,\mathrm{fJ}$~\cite{igzo_energy}, the analog CDS $0.938\,\mathrm{fJ}$, and the P2M unit $1.96\,\mathrm{pJ}$ per OFM. Interconnect latency is dominated by the $1\,\mathrm{GHz}$ ATC ($64\,\mathrm{ns}$); with the $2\,\mathrm{GHz}$ TIA and $\sim40\,\mathrm{GHz}$ optical channel the total is $\approx64.5\,\mathrm{ns}$, versus a $10\,\mu\mathrm{s}$ IPC compute cycle (positive + negative phases). The second layer ($K=3$, stride 1) uses 9 wavelengths transmitting in parallel. In rolling-shutter (RS) mode each row's P2M latency pipelines with the interconnect, so IPC latency dominates at $8.96\,\mathrm{ns}$ for a $560\times560\times3$ image. For $L=10\,\mathrm{mm}$ the total energy is $1.16\,\mu\mathrm{J}$, dominated by the interconnect.

For comparison, we evaluate a digital IPC using electrical interconnects for both RS and global-shutter (GS) modes~\cite{datta2022processing,kaiser2025voltage}; the GS pixel consumes $5.36\,\mathrm{pJ}$/conversion~\cite{kaiser2025voltage} and the electrical interposer $0.26\,\mathrm{pJ/bit}$ at $0.11\,\mathrm{ns}$ for $L=1\,\mathrm{mm}$~\cite{electricalinterposerv1, electricalinterposerv2}. For a fair comparison we let the digital system transmit all ADC bits in parallel, with an 8-bit SAR ADC ($200\,\mathrm{ns}$, $6.4\,\mathrm{pJ}$) and 8-bit DAC ($0.5\,\mathrm{ns}$, $0.24\,\mathrm{pJ}$). Table~\ref{tab:ipc_comp} shows the GS optical configuration achieves the lowest latency and EDP; for RS the interconnect latency is pipelined, so its advantage appears only for $L>30\,\mathrm{mm}$. Fig.~\ref{bar_comp} plots the normalized comparison across link length, calibration, and energy breakdown. The analog IPC (pixel array, IGZO, CDS, ATC) reaches 1.68/118.16 GOPS and 21.61/17.92 TOPS/W for RS/GS.

Precision-matched baseline and distance dependence: The $2.04\times$ EDP improvement in Table~\ref{tab:ipc_comp} is obtained in global-shutter (GS) mode at $L=10\,\mathrm{mm}$ against a column-parallel 8-bit SAR ADC baseline representative of commercial near-sensor readout. This gain is mode-dependent: in rolling-shutter (RS) mode the interconnect latency is pipelined with the P2M compute latency, so the interconnect advantage is masked until $L>30\,\mathrm{mm}$, where the electrical link begins to dominate the overall latency. To isolate the benefit of the time-encoded photonic link from the effect of converter precision, we additionally scale the digital baseline to an iso-precision 6-bit ADC/DAC and sweep the link length (Table~\ref{tab:sixbit}). Even at 6-bit, and without accounting for the serializer/deserializer overhead that a practical bit-parallel electrical link would require, the analog photonic interposer retains an EDP advantage that widens rapidly with distance in GS mode, from $1.31\times$ at $L=10\,\mathrm{mm}$ to $10.07\times$ at $L=25\,\mathrm{mm}$, confirming that the gain originates from the long-distance efficiency of the optical link rather than from a lower converter precision.


\begin{table}[!t]
\begin{center}
\scriptsize\addtolength{\tabcolsep}{-0pt}
\resizebox{\linewidth}{!}{ 
\begin{tabular}{lcccc}
\toprule
\multicolumn{1}{c}{\textbf{Metric}} &
\multicolumn{2}{c}{\textbf{Analog System}} &
\multicolumn{2}{c}{\textbf{Digital System}} \\
\cmidrule(lr){2-3} \cmidrule(lr){4-5}
& \textbf{RS} & \textbf{GS} & \textbf{RS} & \textbf{GS} \\
\midrule
\textbf{Energy ($\mu$J)}        & 1.16 & 1.30 & 1.32 & 1.47 \\
\textbf{Latency (ms)}          & 8.96 & 0.79 & 8.96 & 1.44 \\
\textbf{EDP (nJ$\cdot$s)}       & 10.44 & 1.04 & 11.91 & 2.13 \\
\textbf{Normalized EDP}        & 9.98 & 1 & 11.39 & 2.04 \\
\bottomrule
\end{tabular}
}
\end{center}
\caption{Comparison of the proposed analog IPC system using an photonic interposer and a baseline digital IPC system with electrical interposer, under rolling-shutter (RS) and global-shutter (GS) architectures for L = 10 mm.}

\label{tab:ipc_comp}
\vspace{-0.2 in}
\end{table}

\begin{table}[!t]
\begin{center}
\scriptsize\addtolength{\tabcolsep}{-0pt}
\resizebox{\linewidth}{!}{
\begin{tabular}{lcccc}
\toprule
& \multicolumn{2}{c}{\textbf{Rolling Shutter}} & \multicolumn{2}{c}{\textbf{Global Shutter}} \\
\cmidrule(lr){2-3} \cmidrule(lr){4-5}
\textbf{Link length} & \textbf{Optical} & \textbf{Electrical} & \textbf{Optical} & \textbf{Electrical} \\
\midrule
$L = 10$\,mm & 9.98 & 7.11  & 1 & 1.31  \\
$L = 20$\,mm & 9.98 & 9.40  & 1 & 5.91  \\
$L = 25$\,mm & 9.98 & 10.55 & 1 & 10.07 \\
\bottomrule
\end{tabular}
}
\end{center}
\caption{Normalized EDP of the proposed analog time-encoded photonic interposer versus a \emph{precision-matched} 6-bit electrical digital baseline across link length (normalized to the global-shutter optical case). The optical advantage widens with distance.}
\label{tab:sixbit}
\vspace{-0.2 in}
\end{table}

\subsection{Algorithm analysis with analog non-idealities}

\subsubsection{Non-Linearity and Noise Modeling}

We inject hardware non-idealities and noise into the algorithm: the flow captures the IPC-array non-linearity and the added interposer (TX/RX) non-linearity, and separate Monte Carlo analyses give variations ($\sigma/\mu$) injected into the first layer. The total front-end noise, including transmission offsets, is $\approx14.41\%$. Since the link supports nine wavelengths, the second layer uses a $3\times3$ kernel; deeper layers assume $0.6\%$ noise, consistent with~\cite{10299653}.

\subsubsection{Accuracy Analysis}

We evaluate the end-to-end impact using ResNet-18~\cite{he2016resnet} and MobileNetV2~\cite{sandler2018mobilenetv2} on three workloads of increasing difficulty: VWW~\cite{vww} (binary), CIFAR-10 (multi-class), and ModelNet40 (multi-view 3D recognition, most sensitive to perturbations). Table~2 compares a digital baseline, an IPC-only configuration (analog first layer, no photonic link), and the full IPC + Interposer system.

The baseline input is set to 224×224 so that, after the stride-2 first layer, activations match our 560×560 analog front-end, giving a resolution-independent comparison. The IPC-only case stays within 0.26\% (ResNet-18) and $+0.53\%$ (MobileNetV2) of the baseline on VWW and within a comparable margin on CIFAR-10 and ModelNet40, showing the in-pixel convolution preserves first-layer feature quality under measured non-linearities and noise. Adding the photonic interposer keeps worst-case degradation below 2\% across all workloads (e.g., 1.77\%/1.01\% on VWW). Given the 14.41\% combined interposer variation and 0.6\% accelerator variation, this small penalty, even on the harder multi-class and multi-view tasks, confirms both CNN robustness to analog distortion and generalization beyond VWW.


  \begin{table}[!t]
\begin{center}
\scriptsize\addtolength{\tabcolsep}{-0pt}
\resizebox{\linewidth}{!}{
\begin{tabular}{lcccccc}
\toprule
\multirow{2}{*}{\textbf{Method}} &
\multicolumn{2}{c}{\textbf{VWW}} &
\multicolumn{2}{c}{\textbf{CIFAR-10}} &
\multicolumn{2}{c}{\textbf{ModelNet40}} \\
\cmidrule(lr){2-3} \cmidrule(lr){4-5} \cmidrule(lr){6-7}
& \textbf{R18} & \textbf{MBv2} & \textbf{R18} & \textbf{MBv2} & \textbf{R18} & \textbf{MBv2} \\
\midrule
{Baseline}      & 91.64 & 87.16 & 93.81 & 89.37 & 82.70 & 76.54 \\
{IPC only}      & 91.38 & 87.69 & 93.43 & 89.21 & 82.62 & 75.20 \\
\textbf{Ours}   & 89.87 & 86.15 & 92.92 & 88.90 & 82.29 & 74.84 \\
\bottomrule
\end{tabular}
}
\end{center}
\caption{Accuracy (\%) comparison between the baseline digital inference, the IPC-only implementation, and our proposed approach (IPC + analog photonic interposer) across VWW, CIFAR-10, and ModelNet40 for ResNet18 (R18) and MobileNetV2 (MBv2).}
\vspace{-0.2 in}
\end{table}

\section{Conclusion}

This work presents a time-encoded analog photonic interposer for long-distance, high-fidelity, and energy-efficient transport of analog signals between chiplets. By converting amplitudes into timing intervals using an ATC and transmitting over WDM silicon-photonic links, it embeds an implicit 6-bit quantization within the link and preserves analog information without an explicit high-precision ADC/DAC data-converter pipeline, while requiring only a single wavelength per PE independent of bit precision. We demonstrate its effectiveness in an analog vision pipeline with an in-pixel computing front-end and an analog accelerator, enabling end-to-end analog processing. Simulations in GF 22nm FDSOI and GF 45SPCLO show a $2.04\times$ improvement in energy--delay product over an 8-bit digital electrical link, with the advantage widening with link length even against a precision-matched 6-bit baseline, while maintaining accuracy within 2\% of the digital baseline across VWW, CIFAR-10, and ModelNet40 for ResNet18 and MobileNetV2.

\bibliographystyle{ACM-Reference-Format}

\bibliography{references}

\end{document}